\documentclass[doublecol]{epl2} 
\usepackage{graphicx}
\usepackage{bm}
\usepackage{amsmath}
\usepackage{amssymb}
\usepackage{amsfonts}
\usepackage[colorlinks,
			linkcolor=blue,
            anchorcolor=blue,
            citecolor=blue,
            urlcolor=blue,
            breaklinks=true]{hyperref}

\newcommand{\eref}[1]{Eq.~(\ref{#1})}
\newcommand{\fref}[1]{Fig.~\ref{#1}}
\newcommand{\bra}[1]{\ensuremath{\langle #1|}}
\newcommand{\ket}[1]{\ensuremath{|#1 \rangle}}
\newcommand{\expv}[1]{\ensuremath{\langle #1 \rangle}}
\newcommand{\proj}[2]{\ensuremath{\langle #1 | #2 \rangle}}
\newcommand{\e}{\ensuremath{\varepsilon}}
\newcommand{\p}{\ensuremath{\partial}}
\newcommand{\im}{\ensuremath{\mathrm{i}}}
\newcommand{\df}{\ensuremath{\mathrm{d}}}
\newcommand{\tr}{\ensuremath{\mathrm{tr}}}
\newcommand{\bbR}{\ensuremath{\mathbb{R}}}

\newcommand{\bbI}{\ensuremath{\mathbb{I}}}

\newcommand{\bk}{\ensuremath{\mathbf{k}}}

\newcommand{\calE}{\ensuremath{\mathcal{E}}}

\newcommand{\eg}{\emph{e.g.}}

\title{Corrected Proximity-Force Approximation for Lateral Casimir Forces}
\shorttitle{Corrected Proximity-Force Approximation}
\author{Fanglin Bao\inst{1}\thanks{E-mail: \email{fanglin.bao@coer-scnu.org}} \and Kezhang Shi\inst{1,2}}
\shortauthor{F. Bao \etal}
\institute{                    
  \inst{1} Centre for Optical and Electromagnetic Research, Guangdong Provincial Key Laboratory of Optical Information Materials and Technology, South China Academy of Advanced Optoelectronics, South China Normal University, Guangzhou 510006, China\\
  \inst{2} Centre for Optical and Electromagnetic Research, JORCEP, Zhejiang University, Hangzhou 310058, China
}


\abstract{
The widely-adopted proximity-force approximation (PFA) to estimate normal Casimir forces is known to be asymptotically exact at vanishing separations. In this letter, we propose a correction to the PFA, which is sufficiently accurate in predicting displacement-induced lateral Casimir forces between a sphere and a grating, for separation-to-radius ratio up to 0.5, far beyond the limit within which the application of PFA is previously restricted. Our result allows convenient estimation of Casimir interactions and thus shall be useful in relevant experimental and engineering Casimir applications. We also study the PFA for gradient gratings, and we find that the inhomogeneity-induced lateral Casimir force is beyond the corrected PFA.}

\begin{document}
\maketitle

The Casimir force, resulted from the variation of the zero-point energy when fluctuating electromagnetic (EM) fields are perturbed by materials, is becoming increasingly important in micro- or nano-scale systems \cite{Dalvit2011}. While it is originally predicted between two neutral parallel plates \cite{Casimir1948a}, the Casimir force is usually measured in sphere-plate configurations to avoid the problem of maintaining parallelism \cite{Bressi2002,Mohideen1998,Garrett2018Jan}, where the gently curved sphere is theoretically modelled as a series of patches parallel to the plate and each patch is assumed to interact only with the part of the plate in its close proximity. This so-called proximity-force approximation (PFA) \cite{Blocki1977} relates the normal force between curved surfaces to the Casimir energy per unit area $ \calE $ between corresponding parallel plates via
\begin{equation}\label{eq:PFA0}
F_n=\eta_n\times F_n^{\mathrm{PFA}}, \qquad F_n^{\mathrm{PFA}}=2\pi r \calE,
\end{equation}
and accurately describes the interaction strength in the small-curvature limit, that is, $ \eta_n\to 1 $ when $ \epsilon \equiv d/r\ll 1 $ \cite{Gies2006Jun}, where $ r $ is the radius of the sphere and $ d $ is the surface-to-surface separation of the sphere and the plate. Since contributions of oblique waves to the Casimir force and the non-additivity property of the force are not taken into account directly (but through $ \calE $), the PFA is generally thought of as a rough treatment with uncontrolled errors \cite{Krause2007Jan,Lapas2016Mar,Hartmann2017Jul,Bimonte2017,Garrett2018Jan}, especially for large $ \epsilon $ and for lateral Casimir forces  \cite{Rodrigues2006Mar}. There are evidences, however, that have put \eref{eq:PFA0} on a firm ground. For instance, the semi-classical approximation in Ref.~\cite{Schaden1998Aug,Schaden2000Jan} evaluates the Casimir energy based on classical periodic orbits like in geometrical optics, and recovers \eref{eq:PFA0} within quantum field theory. Refs.~\cite{Bordag2008a,Teo2011Dec} use multiple scattering formalism, confirm \eref{eq:PFA0}, and obtain the first-order correction beyond the PFA, $ \eta_n = 1+\frac{\alpha}{2} \epsilon + \mathcal{O}(\epsilon^2) $. Describing a curved surface by a function $ \psi $ and regarding the Casimir energy as a functional of $ \psi $, \cite{Fosco2011Nov} shows that the PFA coincides with the leading term of the derivative expansion of the Casimir energy, and also obtains the general form of the next-to-leading-order curvature correction. The coefficient $ \alpha $ varies, though, from $ -1.4 $ \cite{Neto2008,Emig2008b} obtained by fitting numeric data, to $ -1.69 $ \cite{Teo2011Dec,Bimonte2012a} and to $ -5.2 $\footnote{Even more complex $ \epsilon\ln\epsilon $ dependence of $ \eta $ is found.} \cite{Bordag2010Mar} obtained by different analytic approaches, for EM fields with perfect-conductor boundary conditions.

\begin{figure}[tb]
\centering
\scalebox{0.39}{\includegraphics{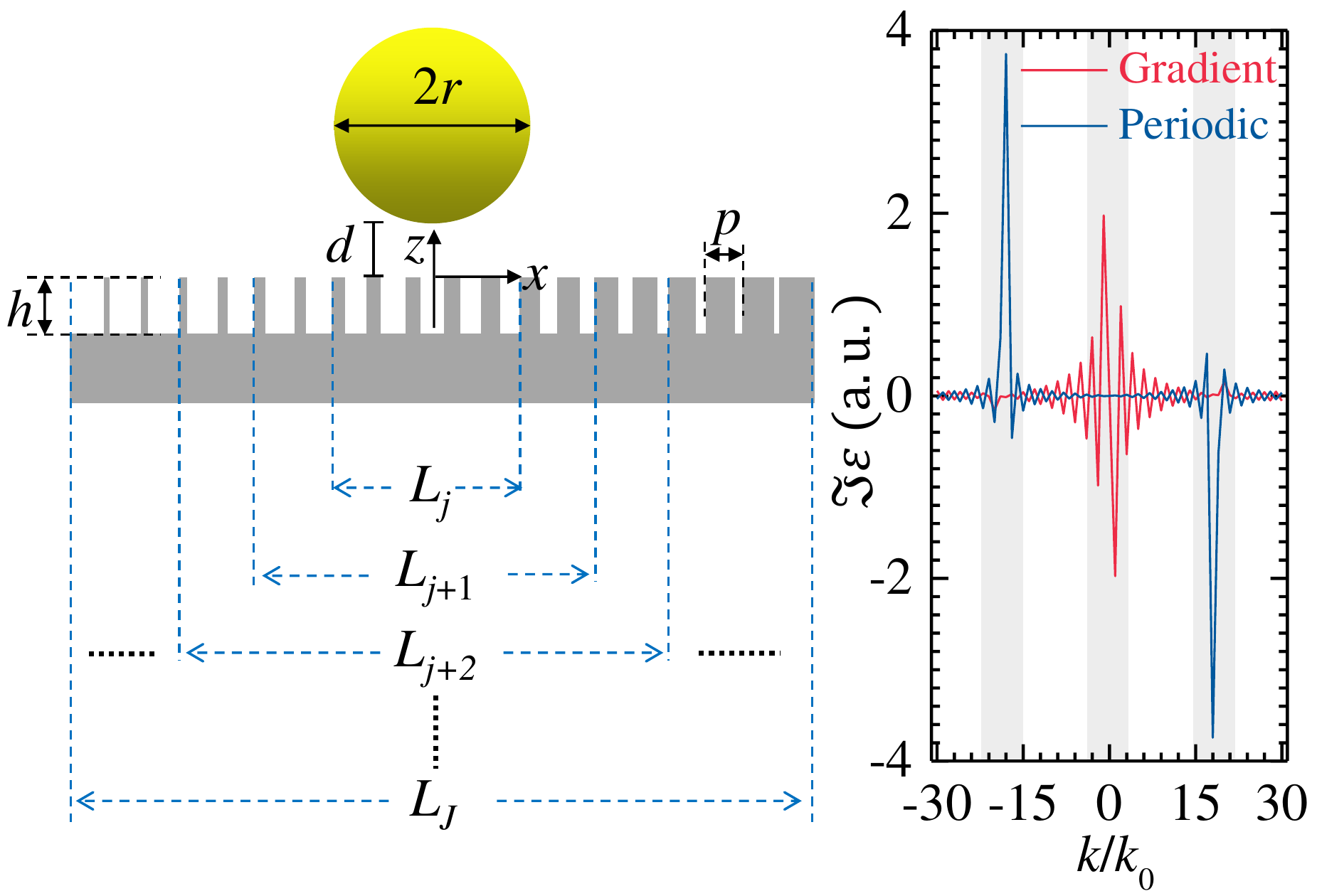}}
\caption{\label{fig:setup} Geometry of the system. A uniform sphere of radius $ r $ is placed at a distance $ d $ above a one-dimensional grating of unit-cell dimension $ p $, super-cell dimension $ L $, and corrugation depth $ h $. Also shown is the schematic of mirror-symmetry breaking of the grating represented by the imaginary part of the $ k $-space permittivity $ \Im\e(k) $. $ k_0=2\pi/L $. }
\end{figure}

Quantifying $ \eta $ allows simple and fast evaluation of Casimir interactions based on the PFA. This is important not only in cases of $ \epsilon<0.1 $ where most experiments are involved and exact calculations become impractical \cite{Bordag2010Mar}, but also in cases of Casimir transport \cite{Bao2018} where lateral Casimir forces ($ F_l\equiv\eta_l\times F_l^{\mathrm{PFA}} $) exerting on a small sphere from a complicated inhomogeneous plate (schematically shown in \fref{fig:setup})
are considered and rigorous computations become cumbersome. Note that $ F_l^{\mathrm{PFA}} $ and $ F_n^{\mathrm{PFA}} $ are related to the approximated Casimir energy $ E^{\mathrm{PFA}} $ via $ F_i^{\mathrm{PFA}}=-\p_i E^{\mathrm{PFA}} $, and $ E^{\mathrm{PFA}}=\int_{\p\Omega} \calE\,\df s $ can be obtained by summing up interaction energy of all patch pairs \cite{Blocki1977}.

In this letter, we report an investigation on $ F_l $ between a sphere and a one-dimensional grating within the $ \epsilon>0.04 $ range, for various filling factors $ f $ ($ f $ is constant for periodic gratings, and varying for ``gradient" gratings) and radius-to-period ratios $ r/p $. We confirm that $ F_l^{\mathrm{PFA}} $ severely deviates from $ F_l $, as previously reported \cite{Rodrigues2006Mar,Messina2015Dec}, but find for periodic gratings that a simple correction $ \eta_l $ exists which precisely recovers $ F_l $ from $ F_l^{\mathrm{PFA}} $ for up to $ \epsilon=0.5 $. The main result of this letter is that: $ \eta_l $ depends on both $ \epsilon $ and $ f $, and reads
\begin{equation}\label{eq:eta}
\eta_l \approx 1.081 -1.215\epsilon -0.122f -0.258\epsilon\cdot f.
\end{equation}
For gradient gratings, apart from the displacement-induced lateral Casimir forces (just like that for periodic gratings, caused by mirror-symmetry breaking corresponding to blue peaks in the right panel of \fref{fig:setup}) there are inhomogeneity-induced lateral Casimir forces (caused by red peaks). The former can be described by \eref{eq:eta} as well, but the latter is beyond the corrected PFA. To support our argument, in what follows we detail the evaluation of lateral Casimir forces and the validity range of \eref{eq:eta}.

The lateral Casimir force in the system under consideration is \cite{Bao2018}
\begin{equation}\label{eq:fl}
F_l=\frac{-1}{\beta}\sideset{}{'}\sum\limits_{n=0}^\infty\sum\limits_{\gamma}\int\limits_{-\infty}^\infty \langle \mathbf{k} \gamma, \mathrm{in} | \bbR_\mathrm{s}\cdot\p_x\bbR_\mathrm{g} | \mathbf{k} \gamma, \mathrm{in} \rangle_n \,\df^2\bk,
\end{equation}
where $ \beta=1/k_BT $, $ k_B $ is the Boltzmann constant, $ T=300\,\mathrm{K} $ is the room temperature, and $ \gamma=\mathrm{TE\ or\ TM} $ represents polarization. The prime on the summation over Matsubara frequencies $ \im\xi_n=\im\cdot 2\pi n/\hbar\beta $ ($ \hbar $ the reduced Planck constant) indicates that the $ n=0 $ term is weighted by $ 1/2 $. $ \bk $ in the plane-wave basis $ | \mathbf{k}\gamma, s \rangle_n $ is the lateral wave vector in the $ x $-$ y $ plane. And $ s=\mathrm{in(out)} $ represents the propagation direction along the negative(positive) $ z $ axis. $ \bbR $ is the reflection operator, and from the translational transformation of $ \bbR $ we have
\begin{equation}\label{eq:momentumEx}
\p_x\bbR_\mathrm{g} = \im\left( \hat{\bk}\bbR_\mathrm{g} - \bbR_\mathrm{g}\hat{\bk} \right).
\end{equation}
\eref{eq:fl} originates from the well-known trace-log formula of Casimir energy for a given system, $ \calE=-\frac{1}{\beta}\sideset{}{'}\sum_n\tr\ln[\bbI-\bbR_\mathrm{s}\bbR_\mathrm{g}] $. The logarithm symbol disappears since in the continuous plane-wave space $ \bbR_\mathrm{s}\bbR_\mathrm{g} $ is infinitesimal ($ \sim\df^2\bk $) and $ \ln[\bbI-\bbR_\mathrm{s}\bbR_\mathrm{g}]=-\bbR_\mathrm{s}\bbR_\mathrm{g} $ exactly holds. The reflection operator of the sphere, observed in plane-wave states, is obtained by partial-wave transformations
\begin{equation}\label{eq:pwtrsfm}
\begin{split}
\bra{\bk\gamma,\mathrm{in}}\bbR_\mathrm{s}
&\ket{\bk'\gamma',\mathrm{out}} =  \frac{1}{2|k_z|K} \sum\limits_{lQ} r_{lQ}\sum_{m=-l}^{l}\\
&\proj{\bk\gamma,\mathrm{out}}{lmQ,\mathrm{in}}\proj{lmQ,\mathrm{in}}{\bk'\gamma',\mathrm{out}},
\end{split}
\end{equation}
where $ K=\xi_n/c $ ($ c $ the speed of light in vacuum), $ |k_z|=\sqrt{K^2+\bk^2} $, $ Q=\mathrm{E(M)} $ in the spherical-wave basis $ \ket{lmQ, \mathrm{s}}_n $ denotes electric(magnetic) multipoles, and $ l $ and $ m $ represent angular momenta. $ r_{lQ} $ is the Mie coefficient, and $ \proj{lmQ,\mathrm{in}}{\bk'\gamma',\mathrm{out}} $ can be found in \cite{Sah2009,Messina2015Dec}. The reflection operator of the periodic grating is obtained by the modal approach (or rigorous coupled-wave analysis) \cite{Davids2010}. With the help of the super-cell technique, modal approach can also be applied to gradient gratings \cite{Bao2018}. For a given gradient grating, we construct a series of virtual metasurfaces (of super-cell dimension $ L_j $, as shown in \fref{fig:setup}) which asymptotically approach the gradient grating. Explicitly, in a $ j_\mathrm{th} $ ($ j=1,2,3,\cdots $) run of computation, we cut a small patch (of dimension $ L_j=j\times p $, nearest to the sphere) of the gradient grating, and define it as a virtual super cell and periodically duplicate it to construct a virtual metasurface, based on which we can calculate the lateral Casimir force $ F_j $ using the modal approach. Then in the $ (j+1)_\mathrm{th} $ run, we enlarge $ L_j $ to $ L_{j+1} $ and get $ F_{j+1} $, and so on. Eventually we can recover the gradient grating when $ L_J\to\infty $, and obtain the desired force $ F $. Since major contributions of the Casimir force exerted on the sphere come from a small area of the grating nearest to the sphere, one can expect that $ F_j $ converges to $ F $ before the $ J_\mathrm{th} $ run.

In computing $ F_l^{\mathrm{PFA}} $, we note that only edges of the grating contribute to the lateral force, since patches of the sphere opposing flat areas of the grating give the same Casimir energy during virtual lateral shifts. Therefore,
\begin{equation}\label{eq:flpfa}
F_l^{\mathrm{PFA}} = \sum_{i}g_i\rho_i\int\limits_{-\pi/2}^{\pi/2} \Big[ 
\calE(z) - \calE(z+h) \Big]\cdot\cos\theta \,\df\theta,
\end{equation}
where $ i $ runs over all edges with coordinates $ x_i^2<r^2 $, $ g_i=1(-1) $ for rising(falling) edges, $ \rho_i=\sqrt{r^2-x_i^2} $, and $ z=d+r-\rho_i\cos\theta $.

\begin{figure}[tb]
\centering
\scalebox{0.92}{\includegraphics{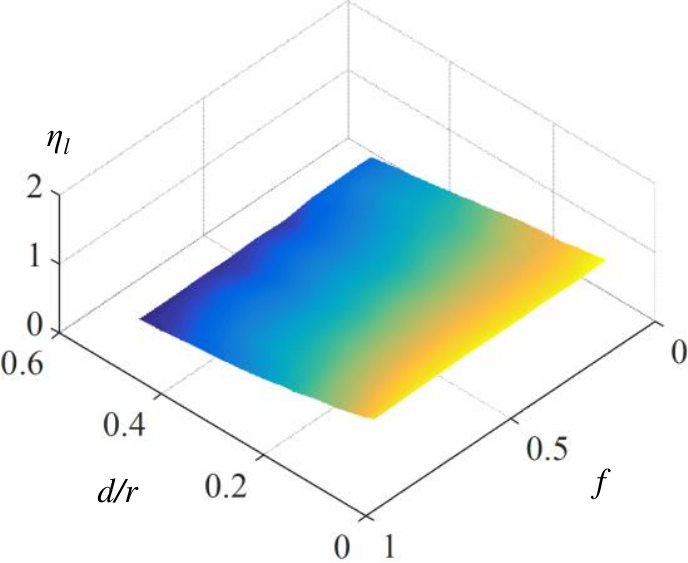}}
\caption{\label{fig:ratioMap} $ \eta_l=F_l/F_l^{\mathrm{PFA}} $ as a function of $ \epsilon=d/r $ and $ f $.}
\end{figure}

For simplicity, we restrict ourselves to $ h=\infty $. This is valid if $ h $ is sufficiently large, due to the decaying nature of Casimir forces. By computing $ F_l/F_l^{\mathrm{PFA}} $ at a rising edge of a periodic grating, for various $ \epsilon $, $ f $ and $ r/p $, we find $ \eta_l $ having a weak dependence on $ r/p $, and almost linearly proportional to $ \epsilon $ and $ f $, as shown in \fref{fig:ratioMap}.
The best fit of \fref{fig:ratioMap} immediately yields \eref{eq:eta}. In the above computations, we have used golden spheres and silica gratings. The Au permittivity is obtained from a Drude model, $ \e_\mathrm{Au}=1+ \Omega^2/\xi_n(\xi_n+\Gamma) $, with plasma frequency $ \Omega=1.28\times10^{16}\,\mathrm{rad/s} $ and damping constant $ \Gamma=6.60\times10^{13}\,\mathrm{rad/s} $. The silica permittivity is fitted by Lorentz terms from tabular data \cite{Palik1985}. In following cases where silicon shall be used, the Drude-Lorentz model is adopted to describe the permittivity with parameters given in \cite{Davids2010}. The highest order of Matsubara frequency is set to be $ n\le\frac{4000}{d[\mathrm{nm}]} $, and the highest order of angular momentum is set to be $ l\le\frac{4}{\epsilon} $, to ensure convergence of $ F_l $. Note that $ \calE=-\frac{1}{\beta}\sideset{}{'}\sum_n\tr\ln[\bbI-\bbR_1\bbR_2] $ can be readily evaluated by the modal approach.

\begin{figure}[htb]
\centering
\scalebox{1}{\includegraphics{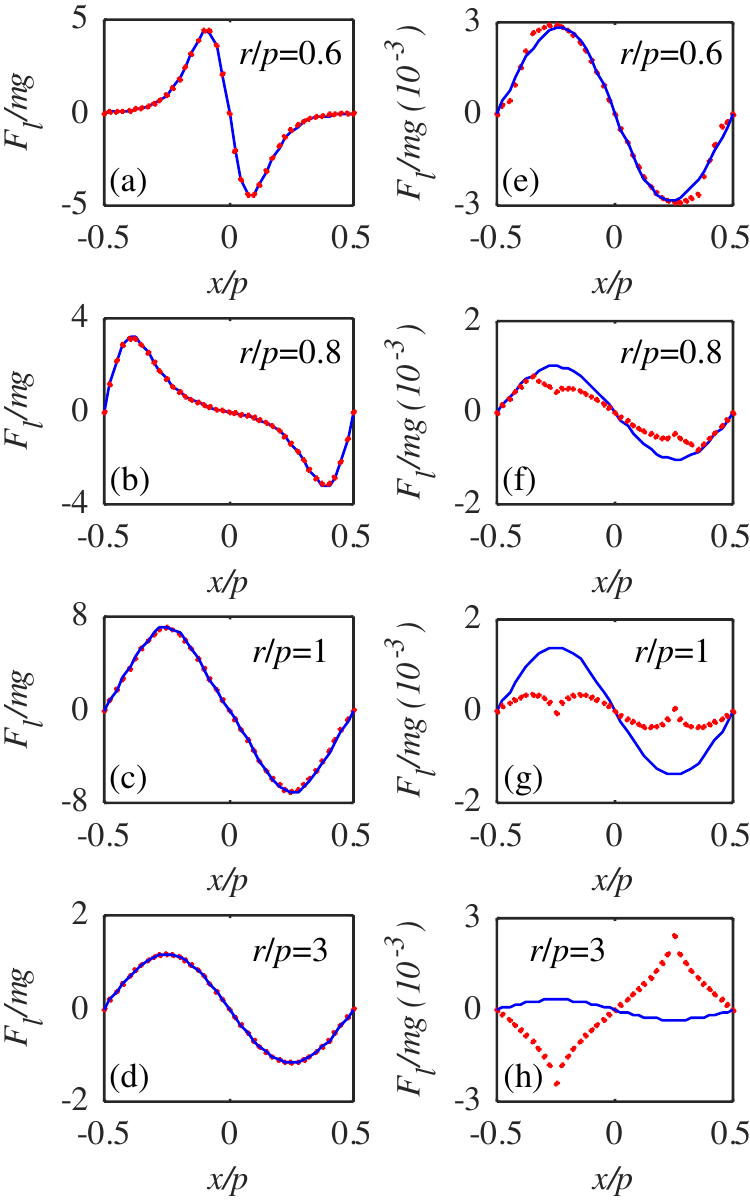}}
\caption{\label{fig:recover} Lateral Casimir forces obtained by rigorous computations (blue curve) and the PFA (red starred), normalized by the weight of the sphere $ mg $. (a,b,c,d): parameters corresponding to blue dots marked in \fref{fig:errorRegion}. (e,f,g,h): red dots in \fref{fig:errorRegion}.}
\end{figure}

\fref{fig:recover} shows the comparison between $ F_l $ and $ \eta_l F_l^{\mathrm{PFA} } $, with $ \eta_l $ given in \eref{eq:eta}. The origin of the $ x $ axis sits at the centre of a ridge of the grating. We can see from Figs.~\ref{fig:recover}(a)-(d) where $ \epsilon=0.04 $, that the corrected PFA precisely recovers $ F_l $ whatever $ f $, $ r/p $, and the force profile are. When $ \epsilon=0.5 $ and $ r/p=0.6 $, the corrected PFA still gives a good estimation of $ F_l $, as shown in \fref{fig:recover}(e). But cases are much worse when $ r/p $ gets larger, as shown in Figs.~\ref{fig:recover}(f-h), even if $ \epsilon=0.18 $ (\fref{fig:recover}(h)), which indicates that the validity condition of the corrected PFA is related to $ r/p $.

\begin{figure*}[h]
\centering
\scalebox{1}{\includegraphics{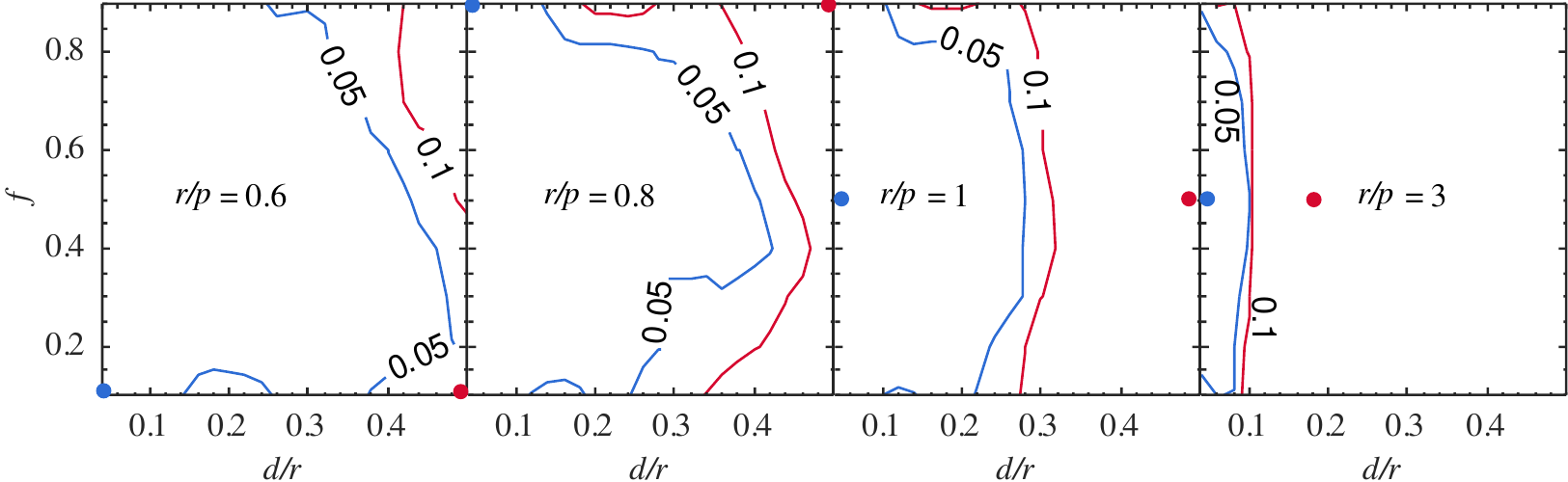}}
\caption{\label{fig:errorRegion} Contour maps of the relative error of the corrected PFA, $ \delta $, as a function of $ \epsilon=d/r $ and $ f $, for various $ r/p $ ($ r=500\,\mathrm{nm} $).}
\end{figure*}

The relative error of the corrected PFA is defined as
\begin{equation}\label{eq:error}
\delta\equiv\frac{\expv{F_l - \eta_l\times F_l^{\mathrm{PFA} }}}{\expv{F_l}},
\end{equation}
where $ \expv{F} $ indicates the integral of $ F^2 $ over a whole period $ p $. The $ 5 \% $ (blue) and $ 10\% $ (red) contours of $ \delta $ are shown in \fref{fig:errorRegion}.
It confirms that \eref{eq:eta} precisely predicts lateral Casimir forces (at the $ 5\% $ level) for up to $ \epsilon=0.5 $ ($ r/p=0.6 $), far beyond the limit within which the PFA is previously restricted ($ \epsilon< 0.00755 $ at the $ 1\% $ level) \cite{Gies2006Jun}. Meanwhile, it can be seen that $ \delta $ modestly increases when $ f\to 1 $ or $ f\to 0 $. Focus on data points corresponding to parameters adopted in \fref{fig:recover}. We can see that four blue dots at $ \epsilon=0.04 $ corresponding to Figs.~\ref{fig:recover}(a)-(d) are indeed in the rather-precise $ \delta<5\% $ region. The red dot for $ r/p=0.6 $ at $ \epsilon=0.5 $ and $ f=0.1 $, corresponding to \fref{fig:recover}(e), is within the $ 5\%<\delta<10\% $ range. Other three red dots for $ r/p\ge 0.8 $ are beyond the $ \delta<10\% $ region, where the PFA is found to yield wrong force profiles even after corrected, as shown in Figs.~\ref{fig:recover}(f)-(h).

\begin{figure}[htb]
\centering
\scalebox{1}{\includegraphics{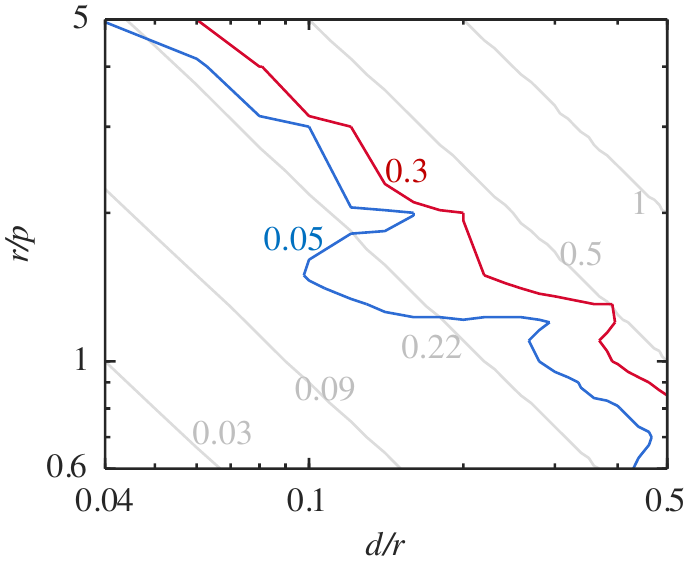}}
\caption{\label{fig:errorCurve} Contour map of the relative error of the corrected PFA, $ \delta $, as a function of $ d/r $ and $ r/p $ ($ r=500\,\mathrm{nm} $, $ f=0.5 $). Only the 0.05 (blue) and 0.3 (red) contours are shown. $ d/p=\mathrm{const.} $ (gray) are also shown as reference lines.}
\end{figure}

The fact that the correction \eref{eq:eta} exists is partially attributed to the geometry of the sphere. The correction actually would not work if the sphere is replaced, for example, with a cube, since in that case the PFA would yield square-wave force profiles while rigorous computations usually yield smoother profiles. In the sphere-grating case, when $ r/p $ increases, more edges contribute to the approximated lateral force and the dominant contribution comes from the edge which is nearest to the sphere. Contributions from other edges are the reason of complex force profiles, and their relative weight, compared with the dominant contribution, increases with $ d $ but decreases with $ p $. $ d/p=r/p\times d/r=\mathrm{const.} $ qualitatively characterizes the contours of $ \delta $, as confirmed by numeric data shown in \fref{fig:errorCurve}, where we can see that \eref{eq:eta} is accurate at the level of $ 5\% $ for about $ d/p<0.22 $. This also explains the shrinking of the validity region of \eref{eq:eta} in \fref{fig:errorRegion}, for increasing $ r/p $.

\begin{figure}[htb]
\centering
\scalebox{1}{\includegraphics{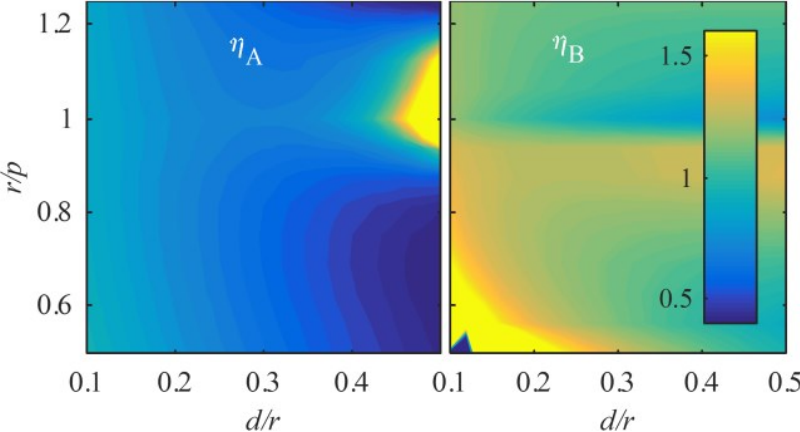}}
\caption{\label{fig:gm} Contour maps of $ \eta_\mathrm{A} $ and $ \eta_\mathrm{B} $, with $ L=4.42\,\mathrm{\mu m} $, $ p=400\,\mathrm{nm} $, and a linear $ f $: $ f(\pm L/2)=0.5\pm 0.25 $. }
\end{figure}

We apply our method to gradient gratings where lateral Casimir forces include displacement-induced and inhomogeneity-induced two components \cite{Bao2018},
\begin{equation}\label{eq:gmfl}
F_l = A\cdot H(x/p) + B,
\end{equation}
where $ H $ is some periodic function of period 1 and of magnitude 1. We decompose both $ F_l $ and $ F_l^\mathrm{PFA} $ as \eref{eq:gmfl}, and study the ratios $ \eta_\mathrm{A}=A/A^{\mathrm{PFA}} $ and $ \eta_\mathrm{B}=B/B^{\mathrm{PFA}} $ accordingly. As shown in \fref{fig:gm}, within the validity region, we find that $ \eta_\mathrm{A} $ is consistent with $ \eta_l $ (errors about $ 10\% $), but $ \eta_\mathrm{B} $ is quite different. In fact, to model the inhomogeneity-induced lateral Casimir force, $ r/p\gg 1 $ should be met so that the information of the gradient of $ f $ can enter into the PFA. This immediately leads to $ d/r\ll 1 $, according to the above analysis, which is obviously outside the $ \epsilon>0.04 $ range under consideration. Therefore, the inhomogeneity-induced lateral Casimir force is beyond the corrected PFA of this letter. Spheres and periodic gratings of other materials, \eg, gold-silicon (sphere-grating) and silica-silica are also studied, and we find that \eref{eq:eta} applies as well, only with previous validity regions shrunk a little.

Our result extends the application of the PFA to previously unexpected region, that is, lateral Casimir forces for $ \epsilon>0.04 $. It allows fast and accurate estimation of Casimir interactions, especially for micro-scale spheres, and therefore our result is important for relevant experimental and engineering Casimir applications concerning dynamics of micro spheres. The method used here also suggests that numeric corrections utilizing spherical geometries might be useful as well, for proximity approximations widely adopted in other branches of physics, for example, to calculate the near-field heat transfer and the nuclear energy.

\acknowledgments
This work was supported by China Postdoctoral Science Foundation (Grant No. 2017M622722).
%


\end{document}